\documentclass[12pt]{article}%
\usepackage{amsmath}
\usepackage{amsfonts}
\usepackage{amssymb}
\usepackage{graphicx}%
\begin{document}
\begin{titlepage}
\begin{flushright}
PUPT-2118\\
hep-th/0405106
\end{flushright}
\vspace{7 mm}
\begin{center}
\huge{Conformal Fixed Points}\par \huge{of
Unidentified Gauge Theories}
\end{center}
\vspace{10 mm}
\begin{center}
{\large
A.M.~Polyakov\\
}
\vspace{3mm}
Joseph Henry Laboratories\\
Princeton University\\
Princeton, New Jersey 08544
\end{center}
\vspace{7mm}
\begin{center}
{\large Abstract}
\end{center}
\noindent
In this article we discuss gauge/string correspondence based
on the non-critical strings
With this goal we present several remarkable sigma models
with the AdS target spaces.
The models have kappa symmetry and
are completely integrable.
The radius of the AdS space is fixed and thus they describe isolated conformal
fixed points of gauge
theories in various dimensions
\vspace{7mm}
\begin{flushleft}
May 2004
\end{flushleft}
\end{titlepage}

Soon after the proposal for gauge/ strings correspondence [1] and its
spectacular implementation in N=4 Yang-Mills theory [2] (which was also based
on the earlier findings [3]) it has been suggested that some
non-supersymmetric gauge theories may become conformal at a fixed coupling
[4,5]. The conjecture was based on the one loop estimate of the $\beta$
function in the $AdS_{p}\otimes S_{q}$ sigma model in the non-critical string
($p+q<10).$The effective equations of motion \ have been shown to have a
solution with a particular values of the radii of $AdS_{p}$ and $S_{q}$ .

Unfortunately this regime takes place at the curvatures of the order of the
string scale where the one loop approximation can be used only as an order of
magnitude estimate. Still, the counting of the parameters in [5] makes this
result plausible. More recently this conjecture was discussed in the zero
dimensional model [ 6].

In this letter I will take the next step and give further arguments that the
above sigma models are conformal . They are also shown to be completely integrable.

The bosonic part of the AdS sigma model is the familiar action for the unit
vector field $\overrightarrow{n}$ which in this case is hyperbolic, satisfying
the relation $\overrightarrow{n}^{2}=-1.$This hyperboloid is embedded in the
$p+1$ dimensional flat space with the signature \ (p,1) or (p-1,2) depending
on whether the dual gauge theory is assumed to be in the Euclidean or
Minkowskian spaces. It is convenient to use the Cartan moving frame, defined
by
\begin{align}
dn  &  =B^{a}e^{a}\\
de^{a}  &  =A^{ab}e^{b}-B^{a}n\nonumber
\end{align}
Here the set of vectors $e_{a},$ $a=1,...p$ are orthogonal to $n.$ The one
forms $B^{a}=A^{a,p+1}$ and $A^{ab}$ form a zero curvature connection with
\ the value in $SO(p+1)$ . Gauge symmetry related to $A^{ab}$ corresponds to
the rotation of the $e-$ vectors by the $SO(p)$ group and thus we treat
$\overrightarrow{n}$ as an element of the coset space $SO(p+1)/SO(p)$ . The
Maurer -Cartan equations (the zero curvature conditions) are%
\begin{align}
dB^{a}  &  =A^{ab}B^{b}\\
dA^{ab}  &  =\frac{1}{2}[AA]^{ab}-B^{a}B^{b}\nonumber
\end{align}
where all the products are the exterior products of 1-forms .

The gauge invariant Lagrangian for the $\overrightarrow{n}$ - field has the
form%
\[
L=\frac{1}{2\gamma}B_{\alpha}^{a}B_{\alpha}^{a}%
\]
where $\gamma$ is a coupling constant and the $SO(p)$ gauge symmetry is
explicit since there are no \ derivatives in this expression. The first
variation of this action is%
\begin{equation}
\delta S\sim%
{\displaystyle\int}
B_{\alpha}^{a}\nabla_{\alpha}\omega^{a}%
\end{equation}
where%
\[
\delta B_{\alpha}^{a}=\nabla_{\alpha}\omega^{a}=\partial_{\alpha}\omega
^{a}-A_{\alpha}^{ab}\omega^{b}%
\]
That gives the equation of motion%
\begin{equation}
\nabla_{\alpha}B_{\alpha}^{a}=0
\end{equation}
In order to calculate the $\beta-$function we have to calculate the second
variation of the action%
\begin{equation}
\delta^{2}S\sim\int\nabla_{\alpha}\omega^{a}\nabla_{\alpha}\omega^{a}-\delta
A_{\alpha}^{ab}\omega^{b}B_{\alpha}^{a}=\int\nabla\omega^{a}\nabla\omega
^{a}+(\omega_{a}\omega_{b}-\omega^{2}\delta_{ab})B^{a}B^{b}%
\end{equation}
where $\delta A^{ab}=\omega^{a}B^{b}-\omega^{b}B^{a}$ is the corresponding
gauge transformation. Using the fact that $<\omega^{a}\omega^{b}>=\frac
{\delta_{ab}}{2\pi}\log\Lambda,$ where $\Lambda$ is a cut-off, we obtain the
divergent counterterm defining the $\beta$ -function%
\begin{equation}
\Delta S=-\frac{p-1}{2\pi}\log\Lambda\int B_{\alpha}^{a}B_{\alpha}^{b}%
\end{equation}
We brought up here this 30 years old derivation because when done this way, it
has a direct generalization for the case of interest. Before coming to that we
need one more recollection - the Wess - Zumino terms and their contribution to
the $\beta$ - function. In the bosonic case the WZ terms exist only for p=3
which corresponds to the coset $\frac{SO(4)}{SO(3)}.$In this case we construct
a 3-form (following Novikov and Witten)%
\begin{equation}
\Omega_{3}=e^{abc}B^{a}B^{b}B^{c}%
\end{equation}
where the exterior product of 1-forms is used.

It is obvious that due to the structure equations this form is closed. It is
also not exact. This last statement requires some explanation. In the compact
case its meaning is obvious - one can't represent $\Omega_{3}=d\Omega_{2}$
with the non-singular $\Omega_{2}$. But what is the meaning of this in the
non- compact and in the supermanifolds ? The definition of cohomology which we
will adopt below is as following. We assume that the 3- form is not exact if
one can't find $\Omega_{2}$ which can be \emph{locally } expressed in terms of
connections. This definition is motivated by the renormalization group, as we
will see below.

The key to the renormlization properties is the variation of $\Omega_{3}%
.$Under the gauge transformation we find
\begin{equation}
\delta\Omega_{3}=e^{abc}d(\omega^{a}B^{b}B^{c})
\end{equation}
Hence%
\begin{equation}
\delta S_{WZ}\sim e^{abc}\int\omega^{a}B^{b}B^{c}d^{2}\xi
\end{equation}
The second variation gives%
\begin{equation}
\delta^{2}S_{WZ}\sim e^{abc}e^{\alpha\beta}\int\omega^{a}\nabla_{\alpha}%
\omega^{b}B_{\beta}^{c}d^{2}\xi
\end{equation}
This term generates $\log\Lambda$ in the second order in $B$ which has an
opposite sign to ( 6). Chosing the action in the form%
\begin{equation}
S=\frac{1}{2\gamma}(\int B^{2}d^{2}\xi+\kappa\int\Omega_{3}%
\end{equation}
we find the the $\beta$- function
\begin{equation}
\beta(\gamma)=\frac{1}{2\pi}\gamma^{2}(1-\kappa^{2})+...
\end{equation}
In the compact case the coefficient $\frac{\kappa}{\gamma}$ must be quantized
and thus (according to the standard argument) can not renormalize. In the
general case we can modify this argument by saying that the counter-terms must
depend on the connection locally, and thus can not create a cohomologically
non-trivial $\Omega_{3}.$There is a nice interplay between the cohomology (in
the sense above) and the renormalization.

In the bosonic case the above construction of the conformal $\overrightarrow
{n}$-field theory is limited to the group $SO(4)$ since there is no invariant
3-tensors in higher dimensional case ( $H^{3}(\frac{SO(p+1)}{SO(p)})=0$ ). In
the superspace the situation is different.

Our main goal is to find the Wess -Zumino terms in the various superspaces of
both critical and non-critical dimensions, which will provide us with the
conformally invariant $\overrightarrow{n}$ -field theories on the world sheet,
corresponding in the hyperbolic cases to various gauge theories in space-time.
The "critical" case of $AdS_{5}\times S_{5}$ has already been examined in the
important work by Metsaev and Tseytlin [7 ]. Our approach in this case leads
to some drastic simplifications, while consistent with their results.

Our first non-trivial example is based on the supergroup $OSp(2\mid4).$Its
bosonic part $Sp(4)\approx SO(3,2)$ acts on $AdS_{4}$ thus describing some 3d
gauge theory. It also has the R-symmetry $SO(2).$This is a simplest choice
because , as we will see below, there is no closed 3-forms without R symmetry
(as in $OSp(1\mid4)$) and the case of the simpler supergroup $OSp(1\mid2),$
which was recently considered in [6 ], is somewhat degenerate and may require
special consideration.

Roughly speaking, the extra invariant tensors in the superspace are simply the
elements of $\gamma-$matrices, while the invariance conditions is given by the
famous $\gamma-\gamma$ identities [ 8]. The set of connections in
$OSp(2\mid4)$ contains as before 1-forms $B^{a}=A^{a5}$ and $A^{ab}$ where the
latter is the gauge connection for $SO(3,1)$ ( $a=1,...4$) ;directions 1 and 5
are assumed to be time-like. This set is complemented with the 2 gravitino
1-forms , $\psi_{i}$, $i=1,2;$each form is also a size 4 Majorana spinor.
Finally we have a connection $C$ of the $R-$ symmetry $SO(2).$ The Maurer -
Cartan equations have the form (they are easily read of the standard
commutation relations of the $OSp$ algebra [9] )%
\begin{align}
dB^{a}  &  =A^{ab}B^{b}+\overline{\psi_{i}}\gamma^{a}\psi_{i}\\
dA^{ab}  &  =\frac{1}{2}[AA]^{ab}-B^{a}B^{b}+\overline{\psi_{i}}\gamma
^{ab}\psi_{i}\\
d\psi_{i}  &  =(\gamma^{ab}A^{ab}+\gamma^{a}B^{a})\psi_{i}+Ce^{ij}\psi_{j}\\
dC  &  =e^{ij}\overline{\psi_{i}}\psi_{j}%
\end{align}
The closed 3-form which replaces (7) in this case is given by%
\begin{equation}
\Omega_{3}=e^{ij}B^{a}\overline{\psi}_{i}\gamma^{a}\gamma^{5}\psi_{j}%
\end{equation}
Here $\gamma^{ab}=\frac{1}{4}[\gamma^{a}\gamma^{b}]$ and $\gamma^{a}$ is the
set of four real gamma matrices; everywhere the antisymmetric product of
differential forms is assumed.

The form $\Omega_{3}$ has explicit gauge symmetry under $SO(3,1)\times SO(2)$
and thus defined on the coset space $\frac{OSp(2\mid4)}{SO(3,1)\times SO(2)}%
.$Let us now calculate $d\Omega_{3}$ by using the above relations. First of
all, due to its explicit gauge symmetry the terms containing $A^{ab}$and $C$
will vanish trivially. The non-trivial part is related to two identities.
First of all , from the $dB$ term comes the contribution%
\begin{equation}
d\Omega_{3}=(\overline{\psi}_{i}\gamma^{a}\psi_{i})e^{kl}(\overline{\psi_{k}%
}\gamma^{a}\gamma^{5}\psi_{l})+...
\end{equation}
It can be rewritten as a sum of terms like
\begin{equation}
(\overline{\psi}_{1}\gamma^{a}\psi_{1})(\overline{\psi}_{1}\gamma^{a}\chi)
\end{equation}
where $\chi=\gamma^{5}\psi_{2}.$Since the product of 1 forms is cyclicly
symmetric, this expression is precisely the $\gamma-\gamma$ identity [8 ] and
is equal to zero. Another dangerous term comes from the pieces $d\psi
=\gamma^{a}B^{a}\psi+...;$and $d\overline{\psi}=-B^{a}\overline{\psi}%
\gamma^{a}$ its contribution is given by%
\begin{equation}
d\Omega_{3}=e^{ij}B^{a}(\overline{\psi}_{i}\gamma^{a}\gamma^{5}d\psi
_{j}-d\overline{\psi}_{i}\gamma^{a}\gamma^{5}\psi_{j})
\end{equation}
(the minus in this formula comes from the fact that we are differentiating
1-forms). By plugging in the above expression for $d\psi$ we see that we get
zero (due to the presence of $\gamma^{5}).$In the simpler case of
$OSp(1\mid4)$ this would not be possible since the expression with $\gamma
^{5}$ is identically zero and without it the contribution (20 ) wouldn't
cancel. Let us also notice that our WZ term is parity -conserving, since the
effect of $\gamma^{5}$ is compensated by the orientation dependence of the
exterior products.

One might think that we found a cohomology but this is not the case. It is
easy to see that $\Omega_{3}=d\Omega_{2},$where $\Omega_{2}=e^{ij}%
\overline{\psi}^{i}\gamma^{5}\psi^{j}.$ The $\gamma-\gamma$ identity turns out
to be a part of the Jacobi identities for $OSp(2|4).$

Now we can chose the action for our sigma model in a remarkably simple form%
\begin{equation}
S=\frac{1}{2\gamma}(\int B_{\alpha}^{a}B_{\alpha}^{a}d^{2}\xi+\kappa\int
\Omega_{2}d^{2}\xi)
\end{equation}
The next step is to find the first and the second variatons of this action. As
in the bosonic case we have to consider the change of $\Omega_{2}$ under
infinitesimal gauge transformations. These transformations are given by%
\begin{align}
\delta B^{a}  &  =\nabla\omega^{a}+\overline{\psi}_{i}\gamma^{a}%
\varepsilon_{i}\\
\delta\psi_{i}  &  =\nabla\varepsilon_{i}+\gamma^{a}B^{a}\varepsilon
_{i}+\gamma^{a}\omega^{a}\psi_{i}%
\end{align}
The bosonic field $\omega^{a}$ and the fermionic fields $\varepsilon_{i}$
(which are two Majorana spinors) will become the degrees of freedom of our
sigma model. The variation of the $\Omega_{2}$is given by
\begin{equation}
\delta\Omega_{2}=e^{ij}(\omega^{a}\overline{\psi}_{i}\gamma^{a}\gamma^{5}%
\psi_{j}+B^{a}\overline{\psi}_{i}\gamma^{a}\gamma^{5}\varepsilon_{j}%
+d(\psi_{i}\gamma^{5}\varepsilon_{j}))
\end{equation}
The last term doesn't contribute to the action and the equations of motion
take the form%
\begin{align}
\nabla_{\alpha}B_{\alpha}^{a}+\kappa e_{\alpha\beta}e^{ij}\overline{\psi
}_{i\alpha}\gamma^{a}\gamma^{5}\psi_{j\beta}  &  =0\\
\gamma^{a}B_{\alpha}^{a}\psi_{i\alpha}+\kappa e_{\alpha\beta}e^{ij}B_{\alpha
}^{a}\gamma^{a}\gamma^{5}\psi_{j\beta}  &  =0
\end{align}
To calculate the $\beta$ - function we need the second variation of the
action. It is sufficient to find it in the background fields for which all the
fermionic components are set to zero. As a result, the answer is a sum of two
terms one of which is quadratic in $\omega$ and given by ( 5), while the other
is quadratic in $\varepsilon.$It is convenient to pass to the complex
notations, $\psi=\psi_{1}+i\psi_{2}$ , and to introduce left and right
components of a spinor, $\psi=\frac{1+\gamma_{5}}{2}\psi_{L}+\frac
{1-\gamma_{5}}{2}\psi_{R}.$ We also set $\kappa=1$ since , as we will see,
this is necessary for conformal symmetry. It is straightforward to vary (21 )
and to find the second variation of the action. In the Weyl notations it is
given by%
\begin{equation}
\delta^{2}S\sim\int(\overline{\varepsilon}_{L}\widehat{B}_{+}\nabla
_{-}\varepsilon_{L}+\overline{\varepsilon}_{R}\widehat{B}_{-}\nabla
_{+}\varepsilon_{R}+\overline{\varepsilon}_{L}\widehat{B}_{+}\widehat{B}%
_{-}\varepsilon_{R})d^{2}\xi
\end{equation}
where $\widehat{B}$ =$\gamma^{a}B^{a}.$

In order to calculate the $\beta-$function it is sufficient to treat the case
in which the background $B$ are constant matrices. This is follows from the
fact that the counterterms can't contain the gradients of $B$ , which would
have higher dimensions. Another constraint on the string-theoretic background
is that the energy- momentum tensors are zero%
\begin{equation}
T_{\pm\pm}=\widehat{B}_{\pm\pm}^{2}=0
\end{equation}
This condition implies that our Lagrangian is degenerate and we must fix the
$\kappa-$ symmetry. In the present context it is quite simple. Redefine the
fields and matrices in the following way%
\begin{align}
\widehat{B}_{\pm}  &  =\sqrt{B_{+}^{a}B_{-}^{a}}\gamma_{\pm}\\
\varepsilon_{L,R}  &  =(B_{+}^{a}B_{-}^{a})^{-\frac{1}{4}}\phi_{L,R}%
\end{align}
where \{$\gamma_{+},\gamma_{-}\}=2;\gamma_{\pm}^{2}=0.$ Just as it is done in
the light cone gauge, we can impose the conditions on $\phi$, which kill one
half of its components. Namely we take $\gamma_{-}\phi_{L}=0;\gamma_{+}%
\phi_{R}=0.$ With these constraints the action (27 ) takes the form%
\begin{equation}
\delta^{2}S\sim\int(\phi_{L}^{\dagger}\nabla_{+}\phi_{L}+\phi_{R}^{\dagger
}\nabla_{\_}\phi_{R}+(\phi_{L}^{\dagger}\phi_{R}+\phi_{R}^{\dagger}\phi
_{L})\sqrt{B_{+}^{a}B_{-}^{a}})d^{2}\xi
\end{equation}
It is instructive to count the number of degrees of freedom in this case. We
see that after fixing the $\kappa-$ symmetry we are left with the two left
movers and two right movers (we are counting real components of the spinors).
On the bosonic side we have four d.o.f. coming from the $AdS_{4}$ which are
reduced to two by the Virasoro constraints. So, there is a match between
bosons and fermions.

The contribution of fermionic fluctuations to the $\beta-$ function comes from
and only from the second order iteration of the mass term
\begin{equation}
\int d^{2}\xi\langle\phi_{L}^{\dagger}\phi_{R}(0)\phi_{R}^{\dagger}\phi
_{L}(\xi)\rangle\sim\log\Lambda
\end{equation}
It has an opposite sign to the ( 6) and cancels it. Beyond one loop we need a
more general argument, since our action is cohomologically trivial. Let us
return to the first variation of the action and write it , using Weyl's
notations, in the form%
\begin{equation}
\delta S=\int(\overline{\psi}_{L-}\widehat{B}_{+}\varepsilon_{L}%
+\overline{\psi}_{R+}\widehat{B}_{-}\varepsilon_{R})d^{2}\xi
\end{equation}
The $\kappa$ -symmetry of this action is immediately seen by setting
$\varepsilon_{L}=\widehat{B}_{+}\kappa_{-}$;$\varepsilon_{R}=\widehat{B}%
_{-}\kappa_{+}$ . We obtain the contribution proportional to the world sheet
energy -momentum tensor $T_{\pm\pm}=(B_{\pm}^{a})^{2}$ which can be cancelled
by the shift of the world-sheet metric. This symmetry will be lost if we
generate a counterterm explicitly dependent on the metric. Thus the non-zero
$\beta$ -function, which introduces an explicit dependence on the Liouville
field , must be forbidden. This, however, is not conclusive since we can't
exclude an anomaly in the $\kappa-$ symmetry. While we lack a complete proof,
let us add another argument in favor of conformal symmetry. The variation of
the action (33 ) doesn't depend on $\psi_{L+}$ and $\psi_{R-}$ . This
independence persists to the second variation. Perhaps one can prove it in all
orders. If this is the case, conformal symmetry follows immediately. Indeed,
the logarithmically divergent counterterm must contain terms like
$\overline{\psi}_{R-}\psi_{L+}$ and thus can't appear. Notice that conformal
symmetry of the familiar WZNW model can be proved by the very similar
argument. However, at present the conformal symmetry is still a conjecture.

It is important to realize that before fixing $\kappa-$ symmetry the model is
not renormalizible. At the first glance it seems strange since both bosonic
and fermionic connection entering (21 ) have dimension one and thus the
coupling constant is dimensionless. On the other hand, even in the flat space
the GS action contains quartic fermionic terms with derivatives which naively
would give power - like divergences. Similar terms appear in our formalism if
we continue the loop expansion by the further variations of the action. The
reason for this discrepancy is that the leading term in the kinetic energy for
the fermionic excitations vanishes. Indeed, while $\psi\sim\partial
\varepsilon$ , the term ($\partial\varepsilon)^{2}$ is absent from the action
due to the properties of the Majorana spinors. Instead we get a kinetic terms
with first derivatives only ( in contrast with the bosonic part). As a result,
the UV dimension of $\varepsilon$ is 1/2 instead of zero. This is the source
of the power-like UV-behaviour. These power-like counter-terms are quite
unusual -by dimensional counting they are seen to contain \emph{negative
powers }of the background field $B.$

After fixing the $\kappa-$ gauge most of the non-linear terms should
disappear. We know that it happens in the light-cone gauge in the flat space
and in the leading UV order the curvature is irrelevant. However, in general
the right choice of the $\kappa$- gauge and renormalizability is a non-trivial
problem. I plan to analyze it in a separate article. Let us stress that
explicit renormalizability may depend on the gauge choice . For example, the
Nambu action of the bosonic string is renormalizible in the conformal gauge
and apparently non-renormalizible in the Monge gauge.

These consideration show that only $\kappa-$ symmetric actions are allowed.
Another reason for that is the fact that in the Minkowskian space - time hte
Green-Schwarz fermions contain negative norms and these are eliminated by the
$\kappa-$ symmetry.

It is interesting to notice that $\kappa-$symmetric models are completely
integrable. In the critical case $AdS_{5}\times S_{5}$ it was known for some
time that this model has a hidden symmetry ( [16 ] and A. Polyakov
(unpublished)). In the non-critical case this is also true and can be
demonstrated in a very simple way. Generally, hidden symmetry follows either
from the Lax representation or from the zero curvature representation with the
spectral parameter $\lambda$ [17 ]. In the latter case we need to construct a
family of $\lambda$ -dependent flat connections, such that at $\lambda=1$ they
coincide with our original set (13-16 ), while the flatness for other
$\lambda$ imply the equations of motion. Let us do it for $OSp(2|4).$ The
relevant zero curvature equations in the complex Weyl notations have the form%
\begin{align}
\nabla_{+}B_{-}^{a}-\nabla_{-}B_{+}^{a}  &  =\overline{\psi}_{L}\gamma^{a}%
\psi_{L}+\overline{\psi}_{R}\gamma^{a}\psi_{R}\\
\nabla_{+}\psi_{L-}-\nabla_{-}\psi_{L+}  &  =\widehat{B}_{+}\psi_{R-}%
-\widehat{B}_{-}\psi_{R+}\\
\nabla_{+}\psi_{R-}-\nabla_{-}\psi_{R+}  &  =\widehat{B}_{+}\psi_{L-}%
-\widehat{B}_{-}\psi_{L+}%
\end{align}
Now, let us introduce the spectral deformation of these connection in the
following way%
\begin{align}
B_{-}  &  \Rightarrow\lambda B_{-};B_{+}\Rightarrow\lambda^{-1}B_{+}\\
\psi_{R\pm}  &  \Rightarrow\lambda^{\frac{1}{2}}\psi_{R\pm};\psi_{L\pm
}\Rightarrow\lambda^{-\frac{1}{2}}\psi_{L\pm}%
\end{align}
while all other connections remain unchanged. These deformations preserve the
zero curvature conditions if the following equations of motion are satisfied%
\begin{align}
\nabla_{+}B_{-}^{a}  &  =\overline{\psi}_{R}\gamma^{a}\psi_{R};\nabla_{-}%
B_{+}^{a}=\overline{\psi}_{L}\gamma^{a}\psi_{L}\\
\widehat{B}_{+}\psi_{L-}  &  =0;\widehat{B}_{-}\psi_{R+}=0
\end{align}
which are just the equations of motion for the $OSP(2|4)$ model. Notice also
that if the fermions are set to zero we get the standard zero curvature
representation for the $\overrightarrow{n}$ -field and the sine -gordon
equations [17 ]. Existence of the $\lambda-$ dependent flat connections easily
leads to the infinite number of conserved currents [17].

In principle with these formulae one can start the heavy machinery of the
inverse scattering method. But even in the bosonic case this is not
straightforward because of the possible quantum anomalies. We will not proceed
with it here and only notice that this hidden symmetry must manifest itself in
the spectrum of the anomalous dimensions.

The above scheme generalizes to the sigma models on $AdS_{5}$ describing 4d
gauge theories. In this case the relevant supergroups are $SU(2,2\mid N)$ .
The bosonic part of it is $SO(4,2)\times U(N)$ (for $N=4$ the right factor is
$SU(4)$ ). It is convenient to use Majorana representation of $SO(4,2)$
provided by the $8\times8$ real $\gamma-$matrices. The conjugation rule in
this case is $\overline{\psi}=\psi^{T}\beta,$ $\widetilde{M}=\beta^{-1}%
M^{T}\beta$ where $\beta$ $=\gamma^{1}\gamma^{6}$ (we assume that $1$ and 6
are the time-like directions). The odd matrices under this conjugation consist
of $\gamma_{pq},\gamma_{pqr}$ $,\gamma_{7}$ (they form the algebra $Sp(8))$ ,
all other tensors are even. The Cartan -Maurer equations are almost the same
as before. We will write explicitly their fermionic part only
\begin{align}
d\psi_{k}  &  =(C_{kl}+i\gamma^{7}D_{kl})\psi_{l}+...\\
dB^{a}  &  =\overline{\psi}_{k}\gamma^{a}\gamma^{6}\psi_{k}+...\\
dA^{ab}  &  =\overline{\psi}_{k}\gamma^{\lbrack a}\gamma^{b]}\psi_{k}+...\\
dC_{kl}  &  =\overline{\psi}_{k}\psi_{l}+...\\
dD_{kl}  &  =\overline{\psi}_{k}\gamma^{7}\psi_{l}-\frac{1}{4}\delta
_{kl}\overline{\psi}_{n}\gamma^{7}\psi_{n}+...
\end{align}
where $C$ is antisymmetric and $D$ is symmetric in $k,l.$ The $U(N)$
connection is just $C+iD.$

Let us discuss now the WZ\ term. Its form depends on the type of the
supercoset space we are looking for, that is on the part of the R-symmetry
group which we want to gauge. This choice must be consistent with the
$\kappa-$symmetry. The general expression for the 2-form defined on $AdS_{5}$
is given by%
\begin{equation}
\Omega_{2}=\overline{\psi}_{k}\gamma^{6}(E^{kl}+i\gamma^{7}F^{kl})\psi_{l}%
\end{equation}
where $E$ and $F$ are some antisymmetric matrices.

Since we do not have a general classification of all possible matrices, let us
discuss some interesting examples. First of all the simplest supergroup is
$SU(2,2\mid1)$ which is the symmetry of the N=1 Yang-Mills theory. In this
case the WZ term doesn't exist , $\Omega_{2}$ vanishes because $\gamma^{6}$ is
an even matrix and the result must be antisymmetric. For the case N=2 we have
a natural action with $E^{kl}=e^{kl}.$ It is easy to see that this
differential form is invariant under the subgroup $SU(2)$ of the $R$ -
symmetry (which is described by the traceless part of the above connection)
and under $SO(4,1)$ transformations of space-time (this symmetry is explicit
in ( 46) ). As a result, the Goldstone modes will as before include bosonic
fluctuations $\omega^{a}$ with $a=1...5,$ two Majorana 8-spinors
$\varepsilon_{k}$ and also the $U(1)$ remainder of the $R-$ symmetry, the
angle $\alpha.$ Thus our action is describing a sigma model on $AdS_{5}\times
S_{1}.$The gauge variations needed to derive the equations of motion are given
by%
\begin{align}
\delta\psi_{k}  &  =\gamma^{b6}(B^{b}\varepsilon_{k}+\omega^{b}\psi
_{k})+e^{kl}\gamma^{7}(C\varepsilon_{l}+\alpha\psi_{l})+\nabla\varepsilon
_{k}\\
\delta B^{a}  &  =\nabla\omega^{a}+\overline{\psi}_{k}\gamma^{a6}%
\varepsilon_{k}\\
\delta C  &  =\nabla\alpha+e^{kl}\overline{\psi}_{k}\gamma^{7}\varepsilon_{l}%
\end{align}
We see that in order to have $\kappa-$ symmetry the action must have the form%
\begin{equation}
S=\frac{1}{2\gamma}\int((B^{a})^{2}+C^{2}+\Omega_{2})d^{2}\xi
\end{equation}
It is convenient at this stage to replace the Majorana 8-spinors by the Weyl
4-spinors. With these modifications the first and the second variations are
the same as in the previous case except that the spinors are larger and the
extra connection $C$ is added. Once again we have a Fermi- Bose match: there
are 6 bosons from $AdS_{5}\times$ $S_{1}$ reduced to 4 by the consraints and 4
physical fermions.

Our next example is the group $SU(2,2\mid4),.$the case already examined in [
7]. As is well known the $R-$ symmetry in this case is reduced to
$SU(4)\approx SO(6).$It is convenient to introduce the Clifford algebra of
$O(6)$ which allows the Majorana representation with the purely imaginary
antisymmetric 8$\times8$ matrices which we will call $\beta^{n}$ , $n=1...6.$
We will now repackage the set of $\psi_{k}$ , $k=1,...4$ connections (each of
which is a Majorana 8-spinor of $SO(4,2)).$We consider a set of 64 Majorana
fields $\Psi$ which are direct product of 8-spinors in $SO(4,2)\times SO(6).$
The Weyl condition, which reduces the number of fields to the desired 32 is
given by $\gamma^{7}\beta^{7}\Psi$=$\Psi.$The set of bosonic connections is
simply doubled. We have to find now the WZ -term. As we saw before, we need an
antisymmetric tensor to write the needed 2-form. The key observation is that
it is provided by the matrix $\beta^{6}$ . The 2-form with the right
properties is
\begin{equation}
\Omega_{2}=\overline{\Psi}\beta^{6}\gamma^{6}\Psi
\end{equation}
This form is explicitly invariant under $SO(4,1)\times SO(5)$ rotations
forming a gauge group. The full action is remarkably simple%
\begin{equation}
S=\frac{1}{2\gamma}\int((B^{a})^{2}+(C^{n})^{2}+\Omega_{2})d^{2}\xi
\end{equation}
The key difference (apart from the different choice of variables) with [ 7] is
that in this paper the WZ term was written as 3-form. Here we notice that this
3-form is exact and this greatly simplifies the matter. Let us also notice
that in [ 7] the authors worked with the pair of the Majorana-Weyl 16-spinors,
$L_{1}$ and $L_{2}.$ In this variables (linearly related to ours ) the form
$\Omega_{2}=\overline{L}_{1}L_{2}$.

We will not repeat the calculations of the second variation and of $\kappa-$
symmetry, since they are practically identical to the derivations given above.
So far we discussed only the $\beta$ -function, but for string theory we also
must have a correct central charge $c(\gamma)=26.$ In principle this relation
determines the value of $\gamma$ and thus fixes the curvature of $AdS_{5}$ .
In the corresponding gauge theory this means that unlike N=4 Yang-Mills theory
we are discussing the 4d theories with the isolated zeroes of their (4d)
$\beta-$ functions. It is clearly important to calculate $c(\gamma).$In the
WZNW model this problem has been solved long ago. In the present case we still
lack the necessary tools. All we can do at the moment is to find this function
at $\gamma\rightarrow0.$In this limit the bosonic part of the action gives a
contribution simply equal to the number of degrees of freeedom. However there
is a subtlety with the fermionic part. The action (27 ) in the UV limit looks
like the action for the world-sheet fermions. The latter have central charge
$\frac{1}{2}$ . So naively one should get $c=\frac{1}{2}($number of
fermi-felds). This counting is wrong (see also related comments in [ 10]). To
get the right one, let us notice that the dependence on the Liouville field in
this Lagrangian appears through the Pauli-Villars regulators. We introduce
heavy fermions $\chi$ with the mass equal to the cut-off $\Lambda.$The mass
term in their Lagrangian has the form%
\begin{equation}
S_{PV}\sim\Lambda\int e^{\varphi}\overline{\chi}\chi d^{2}\xi
\end{equation}
since these fermions are scalars from the world-sheet point of view. For
standard world-sheet fermions, which are spinors we would get $e^{\frac
{\varphi}{2}}$ factor in the corresponding expression. Since the central
charge is the coefficient in front of the Lioville action which is quadratic
in $\varphi,$ we conclude that the right formula for $c$ in the limit of zero
coupling is $c=($ $n_{B}+2n_{F})$ , in which the contribution of the GS
fermions is \emph{four times larger }than the central charge of the
world-sheet fermions. In the case of the flat 10d space that indeed gives
$c=(10+2\times8)=26$ (after the $\kappa-$ symmetry is gauge fixed, we remain
with 8 fermions in each direction).

The sigma models we described, provided that the conjecture of conformal
invariance is correct, describe gauge theories in various dimensions. Some
more work is needed to identified their matter content. In most cases known
today, this issue is resolved by appealing to the D-brane picture in the flat
space and then replacing the D-branes by the corresponding fluxes. This
approach works for the weak coupling when the supergravity approximation is
applicable. However , as was stressed in [ 4] , D-branes, while useful, are
neither necessary nor sufficient for the gauge/strings correspondence.

In general one has to analyze the edge states of the sigma model. As was
argued in [ 1], they are described by the open string vertex operators and
correspond to the various fields on the gauge theory side. Such operators can
be studied at the weak coupling , although even that is non-trivial. These
calculations have not been done so far. The only thing we know at present is
the symmetry of the above models. To avoid confusion one should clearly
distinguish the explicit symmetries of the above actions and the global
symmetry of the theory. The explicit symmetries are in fact gauge symmetries
coming with the coset space. They are related to the right supergroup action.
On the other hand our Lagrangians are written in terms of the left-invariant
connections. Thus the global supergroup $SU(2,2|N)$ of left multiplications is
not visible but definitely present (even in the standard bosonic $\frac
{SO(3)}{SO(2)}$ case the explicit symmetry is $SO(2)$ , while the global
symmetry is $SO(3)$ ).

At the same time there is a simple way to pass to the non-supersymmetric
models. It was pointed out in [ 4] that for the gauge/strings correspondence
it is necessary to eliminate the open string tachion from the edge states. The
minimal way to achieve it in the NSR formalism is to exploit the non-chiral
GSO projection leading to the Type 0 strings without supersymmetry. The closed
string tachyon may be either of the "good variety" [4] in which case it is
harmless or of the bad variety, corresponding to the relevant operators on the
gauge theory side. In the latter case the gauge theory requires a fine-tuning
to be conformal.

In the present context the Type 0 construction in the Green- Schwarz fromalism
corresponds to the summation over the spin structures for the GS fermions (
recall that in the standard supersymmetric case one must take only positive
spin structures). The summation preserves modular invariance and projects out
the states with odd number of GS fermions (see an alternative discussion in
[11 ] ).

Above we discussed only the simplest supercosets. They are cohomologically
trivial and for that reason we couldn't prove non-renormalization of the WZ
term. They also contained no free parameters. It would be very interesting to
find cases without these limitations. A free parameter must appear in the
theories describing gauge fields with the fundamental matter. In this case the
sigma model must contain a parameter $N_{f}/N_{c}$ . It is interesting to
notice that the structure some simple supergroups indeed depends on a free
parameter [9].

Conformal gauge theories described above may find various applications. They
are useful for the further decoding of the gauge/strings correspondence, in
particular for testing of the strong coupling limit which I will discuss
elsewhere. One might also think of using 3d conformal gauge theories for the
holographic description of the early universe. Another interesting problem
related to the above models is QCD with $\vartheta=\pi.$ However, first we
must learn much more about their dynamics (after all we didn't really proved
that the $\beta-$ function is zero and didn't compute the central charge) .

After I wrote this paper I learned (from A. Tseytlin) that the quadratic form
of the WZ term has some history [12-14]. I refer the reader to these valuable
papers. However, neither our models nor the issues of conformal symmetry have
been discussed before. Also, the supercoset models were analyzed in [15 ] in
the Berkovits formalism. Relation of this impressive paper to the present one
is unclear to me.

It is a pleasure to thank Ig. Klebanov, J. Maldacena, A. Tseytlin and H.
Verlinde for very useful discussions.

This work was partially supported by the NSF grant 0243680. Any opinions,
findings and conclusions or recommendations expressed in this material are
those of the authors and do not necessarly reflect the views of the National
Science Foundation.

\bigskip REFERENCES

[1] A. M. Polyakov Nucl. Phys. Proc. Suppl. 68 (1998) ,hep-th/ 9711002

[2] J. Maldacena Adv. Theor. Math. Phys. 2,231 (1998) , hep-th/9711200

[3] Ig. Klebanov Nucl. Phys. B496, 231 (1997)

[4] A. M. Polyakov Int. J. Mod. Phys. A14,645 (1999)

[5] A.M. Polyakov Phys. Atom. Nucl. 64,540 (2001) , hep-th/0006132

[6] H. Verlinde hep-th/0403024

[7] R. Metsaev, A. Tseytlin Nucl. Phys. B533, 109 (1998)

[8] L.Brink, J. Schwarz, J. Sherk Nucl. Phys. B121, 253, (1977)

[9] P. van Nieuwenhuizen, Phys. Reports68,4,(1981), 189

[10] N.Drukker, D. J. Gross, A. Tseytlin JHEP 004 (2000) 021 , hep-th/0001204

[11] Ig. klebanov, A. Tseytlin JHEP 9903 (1999) 015, hep-th/9901101

[12] R. Roiban, W. Siegel JHEP 0011(2000) 024, hep-th/0010104

[13] M. Hatsuda, K. Kamimura, M. Sakaguchi Phys. Rev. D62 (2000) 105024 , hep-th/0007009

[14] M. Hatsuda, M. Sakaguchi hep-th/0205092

[15] N. Berkovits, C. Vafa, E. Witten JHEP 9903 (1999) 018, hep-th/9902098

[16] I. Bena, R. Roiban, J. Polchinski Phys. Rev.D 69, (2004) 046002, hep-th/0305116

[17] B. Dubrovin, I. Krichever, S. Novikov in "Encyclopedia of Mathematical
Sciences" v. 4, Springer Verlag (1990)
\end{document}